\documentclass[a4paper,11pt]{article}
\pdfoutput=1

\usepackage{jcappub}

\usepackage[T1]{fontenc} 

\usepackage{graphicx}
\usepackage{txfonts}
\usepackage{physics}
\usepackage[utf8]{inputenc}
\usepackage[english]{babel}
\usepackage[T1]{fontenc}
\usepackage{lmodern}
\usepackage{microtype}
\usepackage{booktabs}
\usepackage{bm}
\usepackage{graphicx}
\usepackage{amsfonts}
\usepackage{amsmath}
\usepackage{amssymb}
\usepackage{listings}
\usepackage{subfigure}
\usepackage{fancyhdr}
\usepackage{mathtools}
\usepackage{comment}
\usepackage{color}
\newcommand{\review}[1]{#1}

\title{Trust the process: mapping data-driven reconstructions to informed models using stochastic processes}

\author[a]{Stefano Rinaldi}
\author[b,c]{Alexandre Toubiana}
\author[d]{Jonathan R. Gair}

\affiliation[a]{Institut für Theoretische Astrophysik, Zentrum für Astronomie, Universität Heidelberg,\\Albert-Ueberle-Str. 2, 69120, Heidelberg Germany}
\affiliation[b]{Dipartimento di Fisica ``G. Occhialini'', Università di Milano-Bicocca, \\Piazza dell'Ateneo Nuovo, 1, 20126, Milano, Italy}
\affiliation[c]{INFN, Sezione di Milano-Bicocca, Piazza della Scienza 3, 20126 Milano, Italy}
\affiliation[d]{Max Planck Institute for Gravitational Physics (Albert Einstein Institute), Am Mühlenberg 1, 14476, Potsdam, Germany}

\emailAdd{stefano.rinaldi@uni-heidelberg.de}
\emailAdd{alexandre.toubiana@unimib.it}
\emailAdd{jonathan.gair@aei.mpg.de}


\arxivnumber{2506.05153}

\date{\today}

\abstract{Gravitational-wave astronomy has entered a regime where it can extract information about the population properties of the observed binary black holes. The steep increase in the number of detections will offer deeper insights, but it will also significantly raise the computational cost of testing multiple models. To address this challenge, we propose a procedure that first performs a non-parametric (\emph{data-driven}) reconstruction of the underlying distribution, and then remaps these results onto a posterior for the parameters of a parametric (\emph{informed}) model. 
\review{Doing so, instead of performing a full hierarchical analysis per available model, the non-parametric reconstruction can be efficiently remapped onto posterior distributions for each different model.}
In addition to yielding the posterior distribution of the model parameters, this method also provides a measure of the model’s goodness-of-fit, opening for a new quantitative comparison across models.}

\begin{document}

\maketitle

\section{Introduction}
In the ten years since the first detection of gravitational waves (GWs) by the LIGO-Virgo-KAGRA (LVK) collaboration \cite{LIGOScientific:2016aoc}, the field of GW astronomy has evolved from extracting astrophysical information from individual binary black hole (BBH) events to conducting population studies aimed at uncovering the global properties of the observed distributions. The third Gravitational-Wave Transient Catalog (GWTC-3)~\cite{KAGRA:2021vkt} has provided valuable insights into the population of stellar-mass binary black holes (BBHs) in the Universe~\cite{KAGRA:2021duu}, allowing us to infer their mass distribution, to measure the merger rate up to redshift $z \sim 1$, and to begin exploring correlations between parameters at the population level --- all of which carry important information about the formation scenarios of these binaries~\cite{Mapelli:2021taw}.

Population analyses in GW astronomy are typically carried out using a hierarchical Bayesian framework \cite{Mandel:2018mve,Vitale:2020aaz}. Given a model for the source population that depends on a set of parameters --- commonly referred to as \emph{hyperparameters} --- this formalism allows one to infer those hyperparameters while properly accounting for measurement uncertainties and selection effects (namely, the fact that the GW detectors have different sensitivity for sources in different regions of the parameter spaces, e.g., mass and redshift).
Three main approaches have been used to model the source population:
\begin{itemize}
    \item {\bf Astrophysical}: the source population is derived from astrophysical simulations or theory \cite[e.g.,][]{iorio:2023,fragos:2023,briel:2023,fragione:2020,arcasedda:2021,dallamico:2021,romero:2022,torniamenti:2024,tagawa:2020,tagawa:2021,samsing:2022,vaccaro:2024}. These models are straightforward to interpret in terms of physical processes but almost always too computationally expensive to be used in a hierarchical Bayesian framework, \review{thus requiring efficient interpolation techniques to circumvent this issue \cite[e.g.,][]{Zevin:2020gbd,Wong:2020ise,Bouffanais:2020qds,Franciolini:2021tla,Mould:2022ccw,Cheng:2023ddt,Colloms:2025hib,Plunkett:2025mjr}};
    \item {\bf Parametric}: the source population is expressed as a combination of simple, analytically defined functions inspired but not directly linked to the astrophysical processes \cite[e.g.,][]{fishbach:2017,talbot:2018,farah:2023,gennari:2025}. This approach, the most commonly used, provides a certain degree of interpretability for the hyperparameters. At the same time, however, this inference is bound to the specific family of functions chosen for the analysis and can be prone to biases due to mismodelling;
    \item {\bf Non-parametric}: more flexible and complex functional forms capable of approximating arbitrary distributions are used to model the source population, without making strong assumptions on its shape \cite{Tiwari:2020vym,Ruhe:2022ddi,Edelman:2022ydv,Rinaldi,Farah:2023vsc,Callister:2023tgi,Toubiana:2023egi,Heinzel:2024jlc,Mould:2025dts}. Being data-driven, this approach is able to efficiently represent the underlying BBH distribution, but often at the cost of completely losing the physical interpretation of the hyperparameters. 
\end{itemize}
The first two approaches are informed --- albeit at different levels ---  on the physical processes happening in massive stars, whereas the non-parametric approach is completely agnostic in nature.

The growing number of detections in the current and upcoming observing runs will significantly improve our ability to understand the formation of BBHs. However, it will also substantially increase the computational cost of population analyses. This cost arises primarily from the need to account for measurement uncertainties in individual events -- typically handled through Monte Carlo integration -- and from the treatment of selection effects. Importantly, the computational cost increases more than linearly with the number of detections, due to accuracy requirements~\cite{Talbot:2023pex}, making the exploration of multiple population models increasingly impractical. In addition, a quantitative connection between parametric or non-parametric reconstructions and astrophysical models is missing.
Whereas some efficient machine-learning methods are already available \cite{Mould:2025dts} to address the computational cost growth, a direct map between non-parametric and parametric models would open for the use of non-parametric models as a form of \emph{data compression}, greatly reducing the computational complexity of the problem.

In this work, we propose a formalism where the population inference is performed in two stages, similarly to the ideas explored in \cite{Fabbri:2025faf,Ng:2024}. Firstly, a non-parametric method is used to carry out a population analysis that fully accounts for the complexities of hierarchical inference --- such as measurement uncertainties and selection effects --- resulting in a flexible, data-driven representation of the population. This representation is then remapped onto other models --- either parametric or astrophysical --- during the second step, enabling direct comparison among models and interpretation of the data.
Our approach offers a complete statistical framework for performing this remapping and provides a quantitative measure of the goodness-of-fit of the remapped model: this can be viewed as introducing a third hierarchical level to the analysis in which the non-parametric model fitted to the data is treated as a particular realisation of a stochastic process whose expected value is the underlying true population model. In this paper we focus on the Dirichlet processes to describe the remapping but, as we will show, the framework is general and can be applied to any stochastic process.

This paper is organised as follows: in section~\ref{sec:framework}, we describe the general formalism used throughout the paper, specialised to the Dirichlet process case and its practical implementations in section~\ref{sec:dirichlet_process}. The robustness of our approach is then illustrated in two applications, reported in section~\ref{sec:demonstration}: section~\ref{sec:toy_model} analyses the simple case of the inference of a Gaussian distribution, whereas in section~\ref{sec:plp} we demonstrate the performance of our method using an LVK-like astrophysical model. We then conclude with a brief summary of the potential applications in section~\ref{sec:conclusions}.

\section{Framework}\label{sec:framework}
Before presenting the statistical framework developed as part of this work, we start by briefly reviewing the standard formalism for population inference \cite{Mandel:2018mve,Vitale:2020aaz}.

\subsection{Direct inference}\label{sec:std_formalism}
We denote with $\theta$ the set of parameters describing a GW event, $p(d|\theta)$ the single event likelihood and $q_P(\theta|\Gamma)$ the population prior on $\theta$, which depends on hyperparameters $\Gamma$ that we wish to infer. When marginalising over the total rate of events (with a scale invariant prior), the population likelihood for observing $N_{\rm o}$ events $\{d\}=(d_1,...,d_{N_{\rm o}})$ is
\begin{equation}\label{eq:std_no_rate}  
p(\{d\}|\Gamma) = \prod_i^{N_{\rm o}}\int {\rm d}\theta_i \  \frac{p(d_i|\theta_i)q_P(\theta_i|\Gamma)}{p_{\rm det}(\Gamma)}\,. 
\end{equation}
We have introduced the selection function defined as
\begin{equation}
    p_{\rm det}(\Gamma) = \int_{d > {\rm threshold}} \int p(d|\theta) q_P(\theta|\Gamma) {\rm d}\theta \dd d\,,
\end{equation}
where the integral on $d$ is performed over datasets that are considered detectable, in the sense that the chosen detection statistic exceeds a specified threshold, e.g., a false alarm rate smaller than 1/year. 
If we do not wish to marginalise over the total rate $R$, we can use the differential rate $q_R(\theta|\Gamma)$ instead of the population prior $q_P(\theta|\Gamma)$.\footnote{\review{Here we deviate from the standard notation of indicating differential rates as $R(\theta|\Gamma)$. We opted to do this to better highlight that our framework can be applied as-is to both probability densities as well as differential rates.}} The population likelihood then reads
\begin{equation}\label{eq:std_rate}    
p(\{d\}|\Gamma) = e^{-R  p_{\rm det}(\Gamma)}\prod_i^{N_{\rm o}}\int {\rm d}\theta_i \  p(d_i|\theta_i)q_R(\theta_i|\Gamma)\,.
\end{equation}
Note that, by definition, $q_{R}(\theta|\Gamma)$ integrates to $R$, whereas $q_{P}(\theta|\Gamma)$ integrates to 1. 

In both cases, the posterior on $\Gamma$ is obtained by assuming a prior $\pi(\Gamma)$ and using Bayes' theorem:
\begin{equation}
    p(\Gamma|\{d\})=\frac{p(\{d\}|\Gamma)\pi(\Gamma)}{p(\{d\})}\,. 
\end{equation}
The hyperparameters $\Gamma$ enter at the second level of this hierarchical description, while the individual event parameters $\theta$ enter at the first level.

This framework applies both if $q_{P/R}(\Gamma)$ is a parametric model (e.g., \textsc{Power-law+Peak}) as well as if $\Gamma$ describes the potentially infinitely many parameters of a non-parametric model. This second case is of particular interest for this work, due to the lack of direct interpretability of non-parametric models. Instead of using the non-parametric reconstruction to inspire the development of new parametric models --- that will require a new and computationally expensive direct inference per model --- we will now describe how adding a third layer allows for a remapping from $q_{P/R}(\theta|\Gamma)$ (e.g., a non-parametric model) onto another population model $p_{P/R}(\theta|\Lambda)$ (a parametric one) in an efficient way.

\subsection{Remapping at the third hierarchical level}
In the \review{remainder} of this work, we will drop the subscripts $P$ and $R$ as our formalism can be applied in the same way to both the normalised population prior and the differential rate. A remapping from $q(\theta|\Gamma)$ to $p(\theta|\Lambda)$ can be obtained by writing the population likelihood in terms of $\Lambda$ as 
\begin{equation}\label{eq:inital_remapping}
    p(\{d\}|\Lambda,A)=\int p\qty(\{d\}|q(\theta|\Gamma))p\qty(q(\theta|\Gamma)|\Lambda,A) \dd q\,,
\end{equation}
\review{where $A$ denotes the potential additional parameters required by the $\qty(q(\theta|\Gamma)|\Lambda,A)$ term, as discussed below.}
The first term of the integrand is simply the direct inference likelihood where we replaced $\Gamma$ with $q(\theta|\Gamma)$.
We note here that, due to the presence of the stochastic process connecting $q(\theta|\Gamma)$ and $p(\theta|\Lambda)$, the remapped likelihood $p(\{d\}|\Lambda,A)$ is not guaranteed to be the population likelihood $p(\{d\}|\Lambda)$ defined in Eq.~\ref{eq:std_rate} used in the direct inference. In the next section we will show that under specific conditions the two likelihoods are equivalent, but it is worth keeping in mind that this might not be always the case. Given that $q(\theta|\Gamma)$ will be the object of the remapping, we will omit the dependence on $\Gamma$ in the following.

The \emph{conversion} term $p(q|\Lambda,A)$ describes $q$ as a realisation of a stochastic process centred on $p(\theta|\Lambda)$. For instance, a possible choice is a Dirichlet process \cite{ferguson:1973} in the case of normalised distributions, or virtually any probabilistic process for unnormalised distributions (i.e., the differential rate).
In many cases, explicitly writing the probability density of a stochastic process defined on continuous distributions is challenging or even impossible: therefore, we will develop the formalism by first discretising $q$, recovering the continuous distribution limit at a later stage.

The discretisation is achieved by introducing $\bar{q}$ as the histogram built out of a $q$. We decompose it as $\bar{q}=(\bar{B},\bar{Q})$, the binning scheme $\bar{B}$ and the corresponding weights $\bar{Q}$ (counts if the distribution is unnormalised).
With these, Eq.~\eqref{eq:inital_remapping} becomes
\begin{multline}
    p(\{d\}|\Lambda,A)=\int p(\{d\}|q)p(q|\bar{B},\bar{Q}) p(\bar{B},\bar{Q}|\Lambda,A) \dd q \dd \bar{B} \dd\bar{Q}\\
    =\int \frac{p(q|\{d\}) p(\{d\})}{\pi_{1}(q)} \frac{p(\bar{Q}| q,\bar{B})\pi_2(q)}{\pi(\bar{Q}|\bar{B})} p(\bar{Q}|\bar{B},\Lambda,A) \pi(\bar{B})\dd q
 \dd\bar{B} \dd\bar{Q}\,,
\end{multline}
where we used Bayes' theorem twice. The integral over $\bar{B}$ is carried over the space of binning schemes, on which we set a prior $\pi(\bar{B})$. For now, we distinguish between the prior on $q$ used in the first hierarchical inference, $\pi_1(q)$, and the one used in the remapping, $\pi_2(q)$. The term $p(\bar{Q}| q,\bar{B})$ is a Dirac delta, since the weights $\bar{Q}$ are uniquely determined given a binning scheme $\bar{B}$: 
\begin{equation}
    p(\bar{Q}|q,\bar{B}) = \prod_{i=1}^{N_b} \delta\qty(\bar{Q}_i - \int_{\bar{B}_i} q(\theta)\dd\theta)
\end{equation}
The term $\pi(\bar{Q}|\bar{B})$ is then the prior on the bin weights induced by the prior $\pi_2(q)$:
\begin{equation}
    \pi(\bar{Q}|\bar{B}) = \int p(\bar{Q}|q,\bar{B})\pi_2(q)\dd q\,.
\end{equation}

The term $p(\bar{Q}|\bar{B},\Lambda,A)$ is determined by the chosen stochastic process: it measures how likely the chosen model $p(\theta|\Lambda)$ is to generate the bin weights predicted by the model that was first fitted to data, $\bar{Q}(q)$.
Under the assumption that $\pi_1(q) = \pi_2(q)$, we get
\begin{equation}
    p(\{d\}|\Lambda,A)=p(\{d\}) \int \frac{p(q|\{d\})}{\pi(\bar{Q}(q)|\bar{B})} p(\bar{Q}(q)|\bar{B},\Lambda,A)\pi(\bar{B}) {\rm d}q{\rm d}\bar{B}\,.
\end{equation}
The remaining degree of freedom we have is in the choice of $\pi(\bar{B})$, i.e. the binning schemes: in section~\ref{sec:implementation} we will discuss two possible choices. 
The population likelihood can then be evaluated by Monte-Carlo integration using $N_s$ samples of $q$ obtained from the direct inference,\footnote{More precisely, the direct inference yields sample of $\Gamma$, which translates into samples of $q$.} which yields
\begin{equation}\label{eq:mc_dirichlet}
    p(\{d\}|\Lambda,A)\simeq \frac{p(\{d\})}{N_s}\sum_{\substack{q \sim p(q|\{d\}) \\ \bar{B}\sim \pi(\bar{B})}}\frac{ p(\bar{Q}(q)|\bar{B},\Lambda,A)}{p(\bar{Q}(q)|\bar{B})}\,. 
\end{equation}
Assuming a prior choice $\pi(\Lambda,A)$, the posterior on $(\Lambda,A)$ is finally given by Bayes' theorem:
\begin{equation}\label{eq:lambda_a_posterior}
    p(\Lambda,A|\{d\}) =\frac{p(\{d\}|\Lambda,A)\pi(\Lambda,A)}{p(\{d\})}\,
\end{equation}

So far we kept the framework generic, without making any specific choices on distributions or functional forms. In what follows, we will focus on remapping between population distributions --- i.e., normalised probability density functions --- using the Dirichlet processes for the conversion. An alternative derivation assuming a Poisson process can be found in appendix~\ref{app:poisson}.

\subsection{Dirichlet process}\label{sec:dirichlet_process}

The Dirichlet process \review{(DP)}, first introduced in \cite{ferguson:1973}, is a stochastic process defined over the space of probability densities --- meaning that each of its realisations is itself a probability density --- and it is the infinite category limit of the Dirichlet distribution. The Dirichlet distribution is specified for a chosen binning scheme $\bar{B}$ with $N_b$ bins, and is characterised by a set of probability values $\bar{P} = \{\bar{P}_1, \ldots, \bar{P}_{N_{b}}\}$ (the \emph{base distribution}) along with a concentration parameter $\alpha$. The values $\bar{P}$ represent the expected probabilities in each bin, while $\alpha$ controls how tightly the realisations cluster around the base distribution. A large value for $\alpha$ implies lower variance and hence less deviation from $\bar{P}$. 
The functional form of the Dirichlet distribution reads
\begin{equation}
p(\bar{Q}|\bar{P},\alpha)=\frac{\Gamma\qty(\alpha)}{\prod_{i=1}^{N_b} \Gamma\qty(\alpha \bar{P}_i)}\prod_{i=1}^{N_b} \qty(\bar{Q}_i)^{\alpha \bar{P}_i-1}\,,
\end{equation}
where $\Gamma(\cdot)$ denotes the Gamma function.
Here $\bar{P}$ is defined by the probability in each bin predicted by the remapping function $p(\theta|\Lambda)$:
\begin{equation}
    \bar{P}(\Lambda)=\prod_{i=1}^{N_b}\delta\qty(\bar{P}_i-\int_{\bar{B}_i} p(\theta|\Lambda) \dd\theta)\,.
\end{equation}
Thus, we have
\begin{equation}\label{eq:dp_gen}
    p(\bar{Q}|\bar{B},\Lambda,\alpha)=\frac{\Gamma\qty(\alpha)}{\prod_{i=1}^{N_b} \Gamma\qty(\alpha \bar{P}_i(\Lambda))}\prod_{i=1}^{N_b} \qty(\bar{Q}_i(q))^{\alpha \bar{P}_i(\Lambda)-1}\,.
\end{equation}
Let us highlight that this formalism is general and does not depend on the specific choice of binning $\bar{B}$: it can be applied to distributions of any dimensionality, and does not rely on any hypothesis on the shape of the bins, i.e., $\bar{B}$ can be any partition of the parameter space. 
The Dirichlet process is then recovered by taking the infinite number of bins limit. We will now show that this limit gives coherent results for the conversion term.
For the sake of clarity, we will drop the $\Lambda$ and $q$ dependence from $\bar{P}_i$ and $\bar{Q}_i$, respectively.

Assuming that the probability density is defined on a finite interval,\footnote{For distributions that are defined on $\mathbb{R}^n$, like the Gaussian distribution, we can assume that their support is finite thanks to the normalisability requirement and thus constrain them into a finite interval with arbitrary precision.} if the number of bins increases the size of the individual bins decreases, and the probabilities can be approximated as
\begin{align}
    \bar{P}_i \simeq p_i V(\bar{B}_i)\,, \\
    \bar{Q}_i \simeq q_i V(\bar{B}_i)\,,
\end{align}
where $p_i$ and $q_i$ are, respectively, the values of $p(\theta|\Lambda)$ and $q(\theta)$ at the centre of each $\bar{B}_i$, and $V(\bar{B}_i)$ is the volume of each bin. 
For simplicity, we will assume a uniform binning scheme,  $V(\bar{B}_i) = V/N_b$ --- $V$ being the total volume and taken to be $1$ in the following. Introducing the regularised concentration parameter as $\beta\equiv\alpha/N_b$, Eq.~\eqref{eq:dp_gen} can be written as 
\begin{equation}\label{eq:dir_with_beta}
     p(\bar{Q}|\bar{B},\Lambda,\alpha) = \frac{\Gamma\qty(N_b\beta)}{\prod_{i=1}^{N_b} \Gamma\qty(\beta p_i)}\prod_{i=1}^{N_b} \qty(\frac{q_i}{N_b})^{\beta p_i-1}\,.
\end{equation}
The product in the denominator can be expressed as 
\begin{equation}
    \prod_{i=1}^{N_b} \Gamma\qty(\beta p_i) = \exp(\sum_i \ln\qty(\Gamma(\beta p_i))) \simeq \exp(N_b\int \ln\qty(\Gamma(\beta p(\theta|\Lambda)))\dd\theta)\,.
\end{equation}
In the same fashion, 
\begin{equation}
    \prod_{i=1}^{N_b} q_i^{\beta p_i-1} = \exp(\sum_i(\beta p_i -1) \ln(q_i))
    \simeq \exp(N_b \int(\beta p(\theta|\Lambda) - 1)\ln(q(\theta))\dd\theta) \,.
\end{equation}
We can now define
\begin{equation}
\begin{aligned}
    F_1(q,\beta,\Lambda) &\equiv \int\qty(\beta p(\theta|\Lambda) - 1)\ln\qty(q(\theta)) \dd\theta\,,\\
    F_2(\beta,\Lambda) &\equiv \int \ln\qty(\Gamma(\beta p(\theta|\Lambda)))\dd\theta\,.
\end{aligned}
\end{equation}
The product of $1/N_b$ becomes
\begin{equation}
    \prod_{i=1}^{N_b}\qty(\frac{1}{N_b})^{\beta p_i-1} = \qty(\frac{1}{N_b})^{\Sigma_{i=1}^{N_b}\beta p_i-1} = \qty(\frac{1}{N_b})^{N_b(\beta-1)} \,,
\end{equation}
given that $\sum_{i=1}^{N_b} p_i = N_b$ (we recall that the $\bar{P}_i$ are normalised over the bins). As we will show in the following, $\beta$ grows when $q$ is \emph{close enough} to the base distribution $p(\theta|\Lambda)$. In this case, we can use the Stirling approximation for the Gamma function to further develop 
\begin{equation}
    \ln(\Gamma(z)) \simeq \frac{\ln(2\pi)}{2}+\qty(z-\frac{1}{2})\ln(z) - z.
\end{equation}
Applying it to $\Gamma(N_b\beta)$, we get
\begin{equation}\label{eq:closeness}
    p(\bar{Q}|\bar{B},\Lambda,\alpha)\simeq \sqrt{\frac{2\pi}{N_b\beta}}\Big(N_b\ \beta^\beta\ e^{-\beta} e^{F_1(q,\beta,\Lambda) - F_2(\beta,\Lambda)}\Big)^{N_b}\,.
\end{equation}
It is interesting to note that this probability density depends on $q$ and $\Lambda$ only through $F_1(q,\beta,\Lambda)$ and $F_2(\beta,\Lambda)$, and that these functions in turn are exponentiated to the $N_b$-th power. Intuitively, when taking the limit for $N_b\to\infty$, the $\Lambda$ values far from the maximum of $F_1(q,\beta,\Lambda)-F_2(\beta,\Lambda)$ will be exponentially suppressed, eventually producing a Dirac delta-like distribution and effectively mapping every $q$ to a single value of $\Lambda$ corresponding to the $p(\theta|\Lambda)$ that is the \emph{closest} to $q$ using $F_1(q,\beta,\Lambda)-F_2(\beta,\Lambda)$ as distance.

We will now give proof of this intuitive behaviour of $ p(\bar{Q}|\bar{B},\Lambda,\alpha)$. The function $F_2(\beta,\Lambda)$ can also be simplified with the Stirling approximation:\footnote{Recall that we assume the volume $V$ to be 1 and that $p(\theta|\Lambda)$ and $q(\theta)$ are normalised.}
\begin{multline}
    F_2(\beta,\Lambda)\simeq \frac{\ln(2\pi)}{2}-\beta +\beta \ln(\beta) - \frac{\ln(\beta)}{2} \\ +\beta \int p(\theta|\Lambda) \ln(p(\theta|\Lambda)){\rm d} \theta -\frac{1}{2}\int \ln(p(\theta|\Lambda)){\rm d} \theta
\end{multline}
Inserting this in Eq.~\eqref{eq:closeness}, we get
\begin{multline}\label{eq:closeness_simplified}
    p(\bar{Q}|\bar{B},\Lambda,\alpha) \simeq \frac{1}{\sqrt{N_b}}
    \exp \Bigg [ N_b\Bigg ( \frac{N_b-1}{2N_b}(\ln(\beta)-\ln(2\pi))  \\ 
 +\ln(N_b)-\beta D_{\rm KL}(p||q) - \int \ln(q(\theta))\dd\theta + \frac{1}{2} \int \ln(p(\theta|\Lambda)) \dd \theta \Bigg ) \Bigg]\,,
\end{multline}
where we introduced the Kullback-Leibler (KL) divergence \cite{kullback:1951} between $p(\theta|\Lambda)$ and $q(\theta)$:
\begin{equation}
    D_{\rm KL}(p||q) = \int p(\theta|\Lambda) \ln(\frac{p(\theta|\Lambda)}{q(\theta)}) \dd \theta.
\end{equation}
Taking the derivative with respect to $\beta$, we get that the conversion term is maximised for
\begin{equation}
    \beta_{\rm max}=\frac{N_b-1}{2N_b D_{\rm KL}(p||q)}. \label{eq:beta_max}
\end{equation}
So, when $D_{\rm KL}(p||q) \rightarrow 0$, $\beta_{\rm max} \rightarrow \infty$. 
Evaluated at $\beta_{\rm max}$, the conversion term becomes
\begin{multline}\label{eq:conversion_beta_max}
    p(\bar{Q}|\bar{B},\Lambda,\alpha)  \simeq  \frac{1}{\sqrt{N_b}}
    \exp \Bigg [ N_b \Bigg(\frac{N_b-1}{2N_b}\ln(\frac{N_b-1}{4\pi N_b}) + \ln(N_b) -\frac{N_b-1}{2N_b}\ln(D_{\rm KL}(p||q)) \\ -\frac{N_b-1}{2N_b}   - \int \ln(q(\theta)){\rm d} \theta + \frac{1}{2} \int \ln(p(\theta|\Lambda)) {\rm d} \theta\Bigg) \Bigg ]. 
\end{multline}
For sufficiently regular families of probability density functions the integral terms are finite and so, for a fixed $q$, the exponent becomes increasingly large as $D_{\rm KL}(p||q) \rightarrow 0$. This means that, if a $\Lambda_q$ such that $p(\theta|\Lambda_q) = q(\theta)$ exists, the conversion term diverges at $\Lambda_q$. 
Therefore, if the family of models $p(\theta|\Lambda)$ is embedded in the family of models $q(\theta|\Gamma)$ or  $q(\theta|\Gamma)$ is a flexible, non-parametric model that is able to approximate with arbitrary precision certain families of probability distributions $p(\theta|\Lambda)$, the conversion provides an exact mapping. In practice we will have a finite number of samples of $q$, and it is almost impossible that $q(\theta) = p(\theta|\Lambda)$ for any of them unless we choose the same functional form for both families, therefore $\beta_{\rm max}$ will almost always be finite. 

\begin{figure}
    \centering
    \includegraphics[width=0.98\columnwidth]{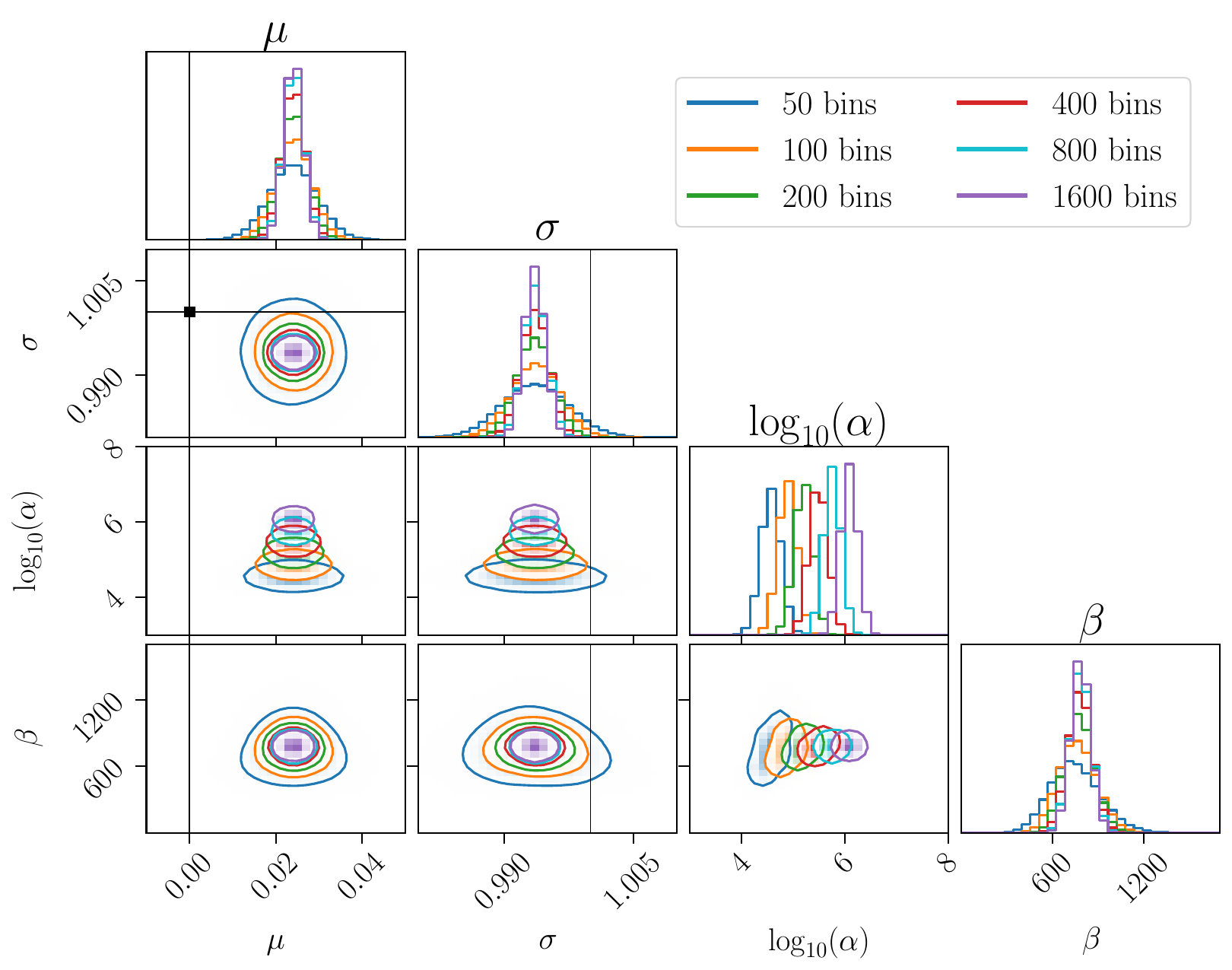}
    \caption{Posterior on $\mu$, $\sigma$, $\log_{10}(\alpha)$ and $\beta$ as a function of the number of bins $N_b$. The contours show the $90\%$ confidence regions and the black \review{cross-hairs indicate} the \review{true} parameters of the Gaussian distribution used to generate data. The posterior on $\alpha$ drifts with the number of bins in such a way that the posterior on $\beta$ remains centred on the same position.}
    \label{fig:comp_nbins}
\end{figure}

Let us return to Eq.~\eqref{eq:closeness}, and investigate the $N_b \rightarrow +\infty$ limit. We define the $\Lambda$, $\beta$ dependent function  appearing in the exponent as\footnote{\review{Rigorously speaking, the function appearing in the exponent is  
 $G_0(\Lambda,\beta,N_b)=\beta \ln(\beta) -\beta + F_1(q,\beta,\Lambda)-F_2(\beta,\Lambda) -
 {\ln(\beta)}/{2N_b}$. However, since from Eq.~\eqref{eq:beta_max}, $\beta$ remains finite, the rightmost term in $G_0$ becomes negligible in the $N_b \rightarrow +\infty$ limit, and $G_0(\Lambda,\beta,N_b)$ becomes independent of $N_b$, reducing to $G(\Lambda,\beta)$.}} 
\begin{equation}
  G(\Lambda,\beta)=\beta \ln(\beta) -\beta + F_1(q,\beta,\Lambda)-F_2(\beta,\Lambda).\label{eq:metric}
\end{equation}
Based on the discussion above, we assume that in the general case $G(\Lambda,\beta)$ admits a maximum at some $(\Lambda_{\rm max}$, $\beta_{\rm max})$. We expand the conversion term around this point as
\begin{multline}\label{eq:expansion}
  p(\bar{Q}|\bar{B},\Lambda,\alpha)\simeq  \sqrt{\frac{2\pi}{N_b\beta_{\rm max}}}\exp \Bigg [N_b \Bigg ( \ln(N_b) +G(\Upsilon_{\rm max}) \\ + \frac{\partial^2G}{\partial \Upsilon_i \partial \Upsilon_j }\Biggr\rvert_{\Upsilon_{\rm max}}(\Upsilon_i-\Upsilon_{{\rm max},i})(\Upsilon_j-\Upsilon_{{\rm max},j}) \Bigg ) \Bigg ]\,,
\end{multline}
where we defined $\Upsilon=(\Lambda,\beta)$. Assuming that $-\qty(\frac{\partial^2G}{\partial \Upsilon_i \partial \Upsilon_j } \Big|_{\Upsilon_{\rm max}}) $ is positive definite, the $\Upsilon$-dependent part is a multivariate Gaussian distribution, with covariance matrix given by 
\begin{equation}
    \Sigma=-\frac{1}{N_b} \left (\frac{\partial^2G}{\partial \Upsilon_i \partial \Upsilon_j } \Biggr\rvert_{\Upsilon_{\rm max}} \right)^{-1}\,.  
\end{equation}
As $N_b\to\infty$, the Gaussian distribution becomes narrower, effectively approaching a Dirac delta distribution. We then get 
\begin{equation}\label{eq:delta_dd}
    p(\bar{Q}|\bar{B},\Lambda,\alpha) \simeq -2\pi \sqrt{\frac{N_b}{\beta_{\rm max}}}  \biggr\rvert\qty( \frac{\partial^2G}{\partial \Upsilon_i \partial \Upsilon_j } \biggr\rvert_{\Upsilon_{\rm max}}) \biggr\rvert ^{-1}
    \exp \Big [N_b \big ( \ln(N_b) +G(\Upsilon_{\rm max}) \big ) \Big ]\delta(\Upsilon-\Upsilon_{\rm max}).
\end{equation}
This behaviour is illustrated in figure~\ref{fig:comp_nbins}. 
In this example, we drew 3,000 samples from a standard Gaussian distribution ($\mu = 0$, $\sigma = 1$) and ran a non-parametric reconstruction on these samples using \textsc{figaro} \cite{rinaldi:2024:figaro} -- a code designed to reconstruct probability densities using a Dirichlet process Gaussian Mixture model, or DPGMM \cite{escobar:1995} -- to draw one non-parametric realisation $q$. We then applied the remapping on this single $q$, assuming a Gaussian distribution for $p(\theta|\Lambda)$. 
Figure \ref{fig:comp_nbins} shows the posterior on $\log_{10}(\alpha)$, $\beta$ and the parameters of the Gaussian distribution for different \review{numbers} of uniform bins. We observe that the posterior becomes narrower and narrower, as expected based on the discussion above. Notice that, while $\alpha$ increases with $N_b$, $\beta$ remains centred around the same value. 

In this framework, the regularised concentration parameter $\beta$ measures the goodness of conversion onto the model $p(\theta|\Lambda)$, growing with the agreement between $p(\theta|\Lambda)$ and $q$ and diverging when the matching between the two functions is perfect (see Eq.~\eqref{eq:beta_max}). Assuming that the non-parametric reconstruction $q$ provides a faithful representation of the data, $\beta$ can be seen as an absolute measure of the goodness of fit of the model $p(\theta|\Lambda)$ to the data. \review{In particular, since from Eq.~\eqref{eq:beta_max} we see that $\beta_\mathrm{max}$ is inversely proportional to the well-understood KL divergence, we expect the value of $\beta$ to have a problem-independent scale when used to quantify how well a parametric model describes the available data (proxied by the non-parametric reconstruction). This self-consistent scale would not necessarily require a comparison between two models as, for example, the Bayes' factor does. We would expect there to be a certain degree of correlation between the Bayes factor of a model relative to a fixed reference model and its $\beta$ value: an extensive study of this relationship is, however, beyond the scope of the current study. Nonetheless, it should be possible to use $\beta$ not only to say which of two models is a {\it better} description of the data, but also to say whether either model is a {\it good} description of the data.}
\review{One caveat to} keep in mind \review{while using the regularised concentration parameter as a model selection tool} is that, since $\beta$ only measures the agreement between the \emph{functions} $p(\theta|\Lambda)$ and $q$, it does not contain a dimensionality penalty factor (often referred to as the \emph{Occam's razor})\review{, which is on the other hand included in the Bayes' factor. Understanding how to account for dimensionality differences while comparing $\beta$ values for different models is another topic that should be investigated in the future.} 

Plugging Eq.~\eqref{eq:delta_dd} for the conversion term into Eq.~\eqref{eq:mc_dirichlet}, we see that the marginalisation over the $q$ samples obtained from the first inference yields a sum of Dirac deltas, with each sample weighted by a quantity capturing the \emph{quality of conversion} (in addition to the prior $\pi(\bar{Q}(q)|\bar{B})$ in the denominator). Samples of $q$ that resemble more closely $p(\theta|\Lambda)$ for some choice of $\Lambda$ have an enhanced contribution that increases with $N_b$. In fact, if we take the ratio of the prefactor in Eq.~\eqref{eq:delta_dd} for two different realisations of $\bar{Q}$, the leading term as $N_b\rightarrow +\infty$ is
\begin{equation}\label{eq:diverging_term}
    \frac{p(\bar{Q}_1|\bar{B},\Lambda,\alpha)}{p(\bar{Q}_2|\bar{B},\Lambda,\alpha)} \propto \sqrt{\frac{\beta_{\rm max,2}}{\beta_{\rm max,1}}}\exp\Big(N_b(G(\Upsilon_{\rm max,1}) - G(\Upsilon_{\rm max,2}))\Big)\,.
\end{equation}
In the following, we will refer to this quantity as the \emph{quality of conversion} factor.
We note that mathematically this is precisely the behaviour we want. Given a sufficiently flexible model, $q$, and infinitely many samples, there will be infinitely many samples in that set which correspond to $p(\theta|\Lambda)$, and these are weighted in the set of samples according to their support in the data. These are exactly the samples we want to extract if we want to recover the posterior that would we obtain with a direct fit to the data. The DP mapping extracts these, and only these, samples, as they are the only samples with $G(\Upsilon)=0$. In practice, however, we will have only a finite number of samples, none of which will exactly correspond to $p(\theta|\Lambda)$. The DP would then pick out only the sample that was closest to being in $p(\theta|\Lambda)$, i.e., the sample with the highest quality of conversion. 
The sum in Eq.~\eqref{eq:mc_dirichlet}, once the large $N_b$ limit is reached, will therefore be dominated by the individual $q$ that is the closest to $p(\theta|\Lambda)$ among all the available $q$ samples. 
Figure~\ref{fig:comp_nbins} illustrates how this can introduce biases in $\Lambda$, since the remapped value of $\Lambda$ corresponding to the dominant $q$ sample may differ from the true value while still lying within the uncertainty expected from a direct inference of $p(\theta|\Lambda)$ using the formalism described in section~\ref{sec:std_formalism}. There are a number of ways that this could be dealt with, and in the following section we will discuss two alternative implementations of our formalism designed to circumvent this issue.

\section{Implementations}\label{sec:implementation}
We present two practical implementations of the Dirichlet process remapping designed to prevent issues with the diverging quality of conversion factor highlighted at the end of the previous section.

\subsection{Unweighted remapping}
One possible strategy to circumvent the issue pointed out at the end of the previous section is to treat each $q$ sample drawn from $p(q|\{d\})$ separately, making use of the fact that we can map each of them to a single $(\Lambda,\beta)$ via the infinite bins limit. We can do this either for the whole set of samples, or first select only those samples that are ``sufficiently close'' to the target distribution, by setting a threshold on the quality of conversion factor.

For a large but finite number of bins, Eq.~\eqref{eq:dir_with_beta} defines a non-singular probability density for $(\Lambda,\beta)$ that can be explored with a stochastic sampler or with a maximisation algorithm. In the previous section we have shown that, for a given $q$, the point $(\Lambda_{\rm max},\beta_{\rm max})$ is independent of the number of bins once the assumption of large $N_b$ is met: this means that the maximum of the finite-bins distribution will coincide with the position of the delta-like distribution obtained taking $N_b\to \infty$. 
Instead of using the likelihood in Eq.~\eqref{eq:mc_dirichlet} that involves a Monte Carlo sum over diverging terms, with this approach we map every $q$ sample to its corresponding $(\Lambda_{\rm max},\beta_{\rm max})$ point: these samples are then weighted according to the prior $\pi(\Lambda,\beta)$ of Eq.~\eqref{eq:lambda_a_posterior} to obtain samples from $p(\Lambda, \beta| \{d\})$.

Operatively, this approach can be implemented as follows: a single $q$ sample is drawn using a non-parametric method from $p(q|\{d\})$ and then discretised using a binning scheme $\bar{B}$ with a large enough but finite $N_b$. In our investigations we found that usually $N_b \gtrsim \mathcal{O}(40)$ is already large enough, and the results we got did not improve significantly with a larger $N_b$.
We then use a maximisation algorithm to locate the $(\Lambda,\beta)$ point that maximises
\begin{equation}
    p(\bar{Q}|\bar{B},\Lambda,\beta) \times \frac{\pi(\Lambda, \beta)}{p(\bar{Q}|\bar{B})}
\end{equation}
using the one $q$ sample mentioned above: this will be our corresponding $(\Lambda,\beta)$ sample. Repeating this procedure for multiple $q$ realisations will yield a set of posterior samples for $p(\Lambda,\beta|\{d\})$. As we are not including the quality of conversion factor among different samples, this approach will yield different posteriors on $\Lambda$ than the direct inference. This is because the samples are weighted in the fitted posterior according to how well the corresponding $q$ fit the data, not the $p(\theta|\Lambda)$ that best-matches that $q$. However, this procedure is likely to be conservative in that we expect the posteriors to be broader because they include samples that fit the observations less well.

In this work, as a non-parametric method we use \textsc{figaro} \cite{rinaldi:2024:figaro}, a Python code based on the DPGMM,\footnote{Publicly available at \url{https://github.com/sterinaldi/figaro} and via \texttt{pip}.} but this remapping scheme can be applied to every non-parametric method. This specific implementation encodes a uniform prior on $q$, therefore $p(\bar{Q}|\bar{B})$ becomes the symmetric Dirichlet distribution on the $N_b$-dimensional simplex:
\begin{equation}
    p(\bar{Q}|\bar{B}) = \Gamma(N_b)\,. 
\end{equation}
The minimisation algorithm we use is the Dual Annealing global optimiser provided by \texttt{Scipy} \cite{scipy:2020} (\texttt{scipy.optimise.dual\_annealing}). The infrastructure we developed is publicly available and can be found at \url{https://github.com/sterinaldi/NP2P}.

\subsection{Flexible binning}
In the second implementation, the non-parametric reconstruction $q$ is a binned histogram, where both the number and positions of the bins are free to vary thanks to the use of reversible-jump Markov Chain Monte Carlo (RJMCMC). We use the RJMCMC implementation of the \texttt{Eryn} sampler\footnote{Publicly available at \url{https://github.com/mikekatz04/Eryn}.}~\cite{Karnesis:2023ras}. In this case, the non-parametric reconstruction already provides a binning scheme $\bar{B}_q$, and for the prior on the binning schemes used in the remapping, we adopt the binning scheme \review{that} is learned from the data. The binned histogram is normalised to the total rate; that is, we use Eq.~\eqref{eq:std_rate} for the first inference and then renormalise a posteriori. The likelihood is computed using Eq.\eqref{eq:mc_dirichlet}, with the Dirichlet distribution (Eq.~\eqref{eq:dp_gen}) applied to the conversion term. Assuming a flat prior between $0$ and a fixed maximum for the bin counts in the non-parametric reconstruction, the resulting prior on the normalised bin counts is $p(\bar{Q}(q)|\bar{B}) = N_b^{N_b - 1}$. Finally, we assume a log-flat prior on the regularised concentration parameter $\beta$. The prior on the hyperparameters $\Lambda$ is model dependent.

Even for a finite and reasonably small number of bins—typically $N_b \sim \mathcal{O}(10)$ in the examples considered in this paper—we find that the sum over non-parametric samples in Eq.~\eqref{eq:mc_dirichlet} can be dominated by a small subset of samples. This is due to the quality of conversion factor, which often results in an effective sample size of only $\sim 100$ out of $10^{4}$ total samples.
To mitigate this effect, we replace the mean computed in Eq.~\eqref{eq:mc_dirichlet} with the median over the non-parametric reconstructions. This can be justified by the central limit theorem, which states that the Monte-Carlo sum should be normally distributed around the theoretical value of the integral. For a Gaussian distribution, the mean and median coincide, but the median is more robust to outliers, which in this context are the few samples with larger weights.
This modification reduces the inferred values of the regularised concentration parameter $\beta$, since it reduces the importance of the non-parametric reconstructions $q$ that happen to convert particularly well to some $p$ and would thus yield larger $\beta$ values: at the same time, however, this procedure returns more robust estimates for the hyperparameters $\Lambda$. 

We want to highlight that the key aspect of this implementation is that the binning scheme is learned from the data and, therefore, provides a natural choice for the binning scheme(s) used in the remapping. Other non-parametric reconstructions relying on a partitioning of the parameter space --- not necessarily histograms --- could also be used. 

\section{Applications}\label{sec:demonstration}
After introducing the statistical framework and briefly summarising our implementations, we now apply this method to two simulated examples to demonstrate its robustness. The functional form of all the distributions used in this section can be found in appendix~\ref{app:pop_model} together with the prior intervals that we used in the analysis.

\subsection{Gaussian distribution}\label{sec:toy_model}
The first example we present is a standard Gaussian distribution ($\mu = 0$, $\sigma = 1$), and apply the two implementations of our formalism to a dataset obtained by drawing samples from this distribution assuming a uniform prior on $\mu$ and $\sigma$ in the remapping. For comparison, we also analysed these data using the direct inference method (section~\ref{sec:std_formalism}).

\begin{figure}
    \centering
    \includegraphics[width=0.98\columnwidth]{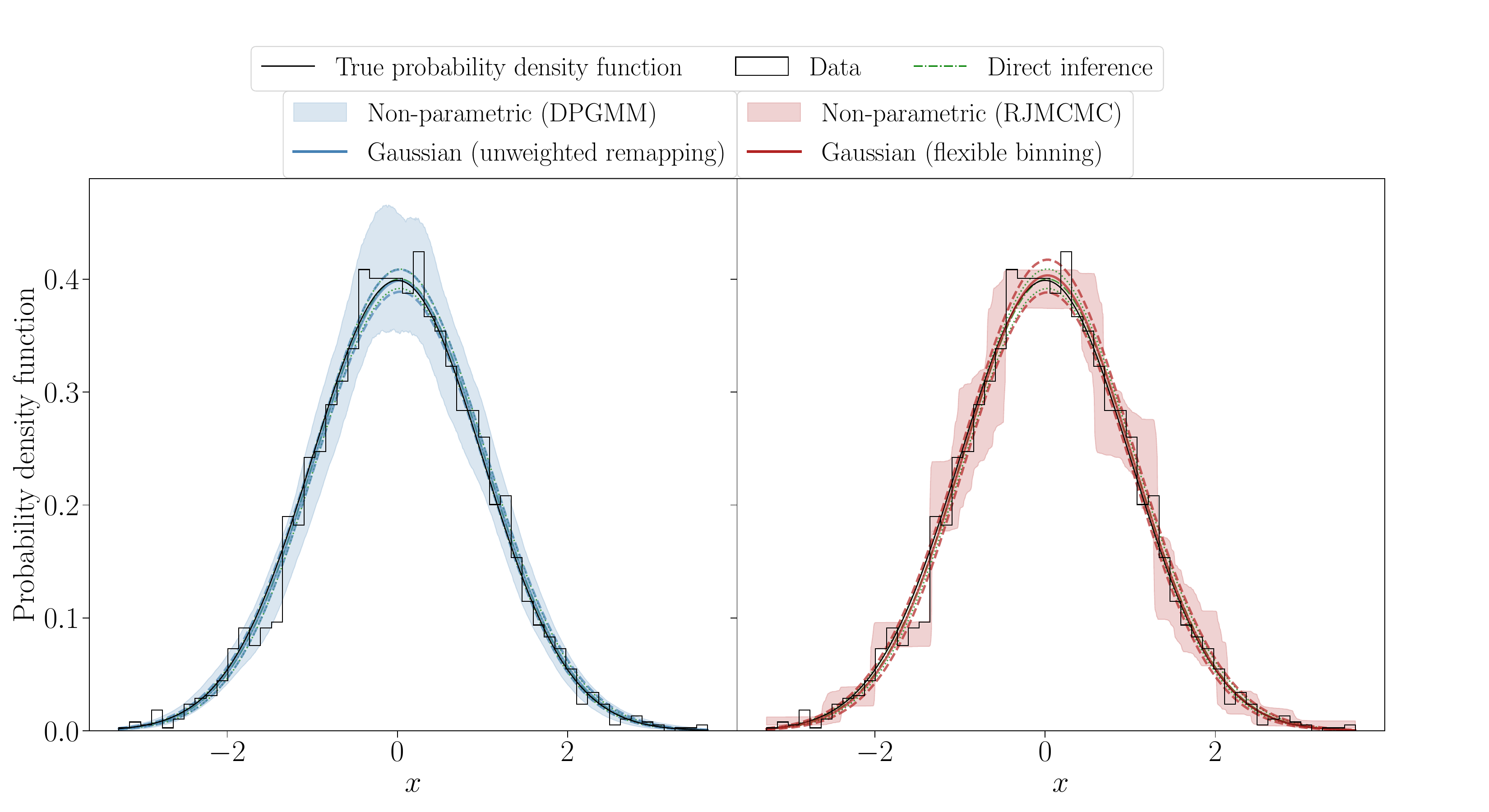}
    \caption{Comparison between the non-parametric reconstructions (shaded areas, 90\% credible region), the remapped Gaussian distributions (dashed lines, median and 90\% credible region) and the true probability density function (solid black line), along with the histogram of the simulated data and the result obtained by direct inference (dot-dashed green lines, median and 90\% credible region). Left panel refers to the unweighted remapping approach, right panel to the flexible binning method.
    }
    \label{fig:comp_ppds}
\end{figure}

\begin{figure}
    \centering
    \subfigure[Gaussian distribution]{
    \includegraphics[width=0.47\columnwidth]{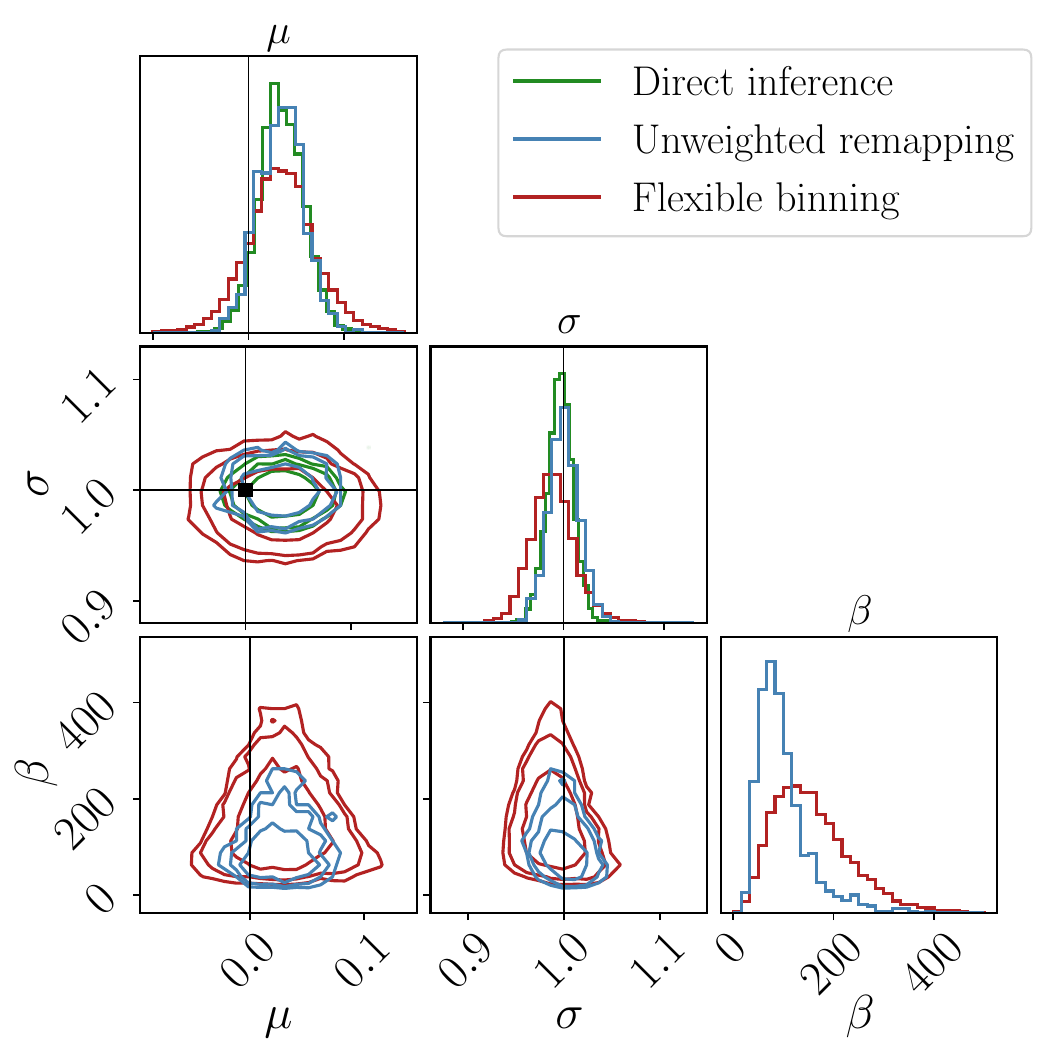}
    \label{fig:comp_post}
    }
    \subfigure[Cauchy distribution]{
    \includegraphics[width=0.47\columnwidth]{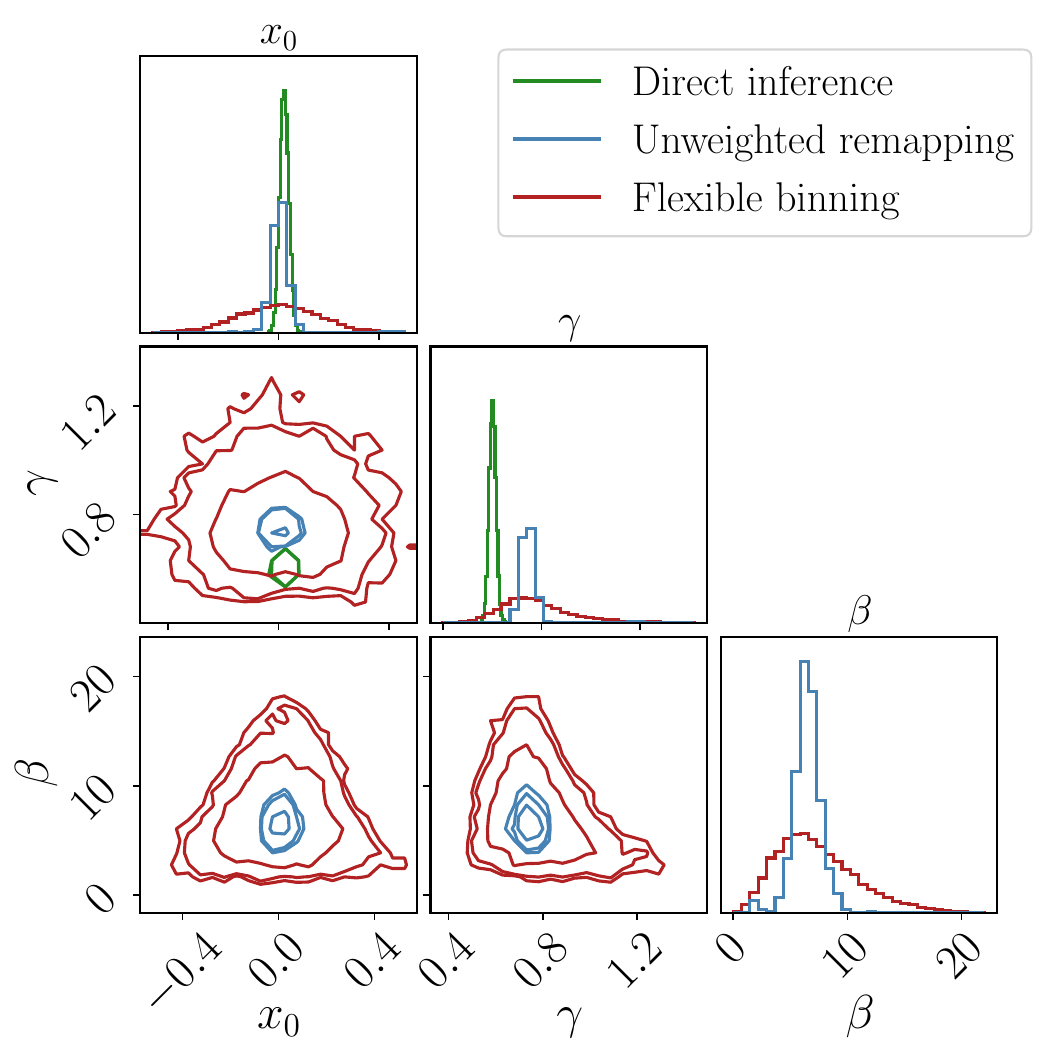}
    \label{fig:comp_cauchy}
    }
    \caption{Posterior distributions obtained via remapping onto a Gaussian (left) and Cauchy (right) distribution using the two approaches described in this work and via direct inference. The contours show the $68$, $90$ and $95\%$ confidence regions and the black \review{cross-hairs} mark the true values of $\mu$ and $\sigma$ for the Gaussian case.}
\end{figure}

In figure~\ref{fig:comp_ppds} we report the probability density function used to generate the data, the simulated dataset (3,000 samples), and the two non-parametric reconstructions --- one using \textsc{figaro}, and the other based on the flexible binning approach --- along with the inferred parametric distributions.
Figure~\ref{fig:comp_post} presents the posterior distributions on $\Lambda = (\mu,\sigma)$ and $\beta$ obtained by remapping the non-parametric reconstructions onto a Gaussian distribution. For both approaches, the true values lie within the posterior support, and the resulting distributions are consistent with those obtained via direct inference.
\review{The remapping procedure took $o(1)$ minute for both approaches. Please note that this is the elapsed time for the remapping procedure only and does not include the time taken to build the non-parametric reconstruction in the first instance.}

The flexible binning approach yields broader posteriors than the other two approaches. This is not surprising: even if the remapped model $p(\theta|\Lambda)$ and the non-parametric reconstruction belonged to the same family of distributions, the number of bins used in the flexible approach is too small for the conversion term to approximate the weighted delta function in Eq.~\eqref{eq:delta_dd}. In this situation, each $q$ induces a posterior on $\Lambda$ with a finite width, rather than a single point estimate, and the resulting posterior distribution is given by their weighted superposition, where the weights are determined by the quality of conversion factor.
The unweighted remapping posteriors are also slightly broader than those from direct inference, due to not including the quality of conversion factor.  The confidence intervals for the posterior predictive distributions reported in figure~\ref{fig:comp_ppds} --- obtained using the posteriors on $(\mu, \sigma)$ obtained via the remapping procedure --- are consistent with the direct inference posterior predictive distribution for both implementations, as well as encompassing the true probability density function. \review{With this simple example we showed that our method is consistent with expectations: a more thorough assessment of the statistical robustness of the method via a pp-plot test is deferred to Section~\ref{sec:plp_prior}.}

In a real-world scenario we would have at hand a variety of potential models to describe the available data: therefore, we repeated the exercise of remapping our non-parametric reconstruction to four other parametric models: a generalised Gaussian distribution, a Cauchy distribution, an exponential distribution and a uniform distribution. The details of these models can be found in appendix~\ref{app:pop_model}. In figure~\ref{fig:comp_beta}, we compare the posteriors on $\beta$ obtained by remapping onto these different families of distributions and rank them accordingly.
The highest values of $\beta$ are found for the two models that encompass the true (simulated) distribution --- namely, the Gaussian and the generalised Gaussian -- recalling that $\beta$ does not include a dimensionality penalty.
The other distribution families are disfavoured with respect to these two: the Cauchy and exponential distribution, despite their bell-like shape, have tails that do not match the simulated Gaussian distribution, whereas the uniform distribution displays none of the features found by the non-parametric inference and thus gets heavily suppressed.

In figure~\ref{fig:comp_cauchy} we report the posterior distribution on the parameters of the Cauchy distribution obtained via both remapping approaches as well as via direct inference.
We observe that the direct inference and the unweighted remapping posteriors exhibit little overlap, particularly for the $\gamma$ parameter. The flexible binning posterior encompasses both distributions, but with a maximum a posteriori value more in agreement with the weighted remapping posterior. This demonstrates that, for models that provide a poor description of the data, neither remapping procedure provides the same posterior as direct inference, although the flexible binning procedure can still provide a valid parameter estimate, as it inflates the posterior uncertainty.

This discrepancy arises because the remapped posterior distribution is conditioned on the initial non-parametric reconstruction: in other words, the remapping procedure identifies the parameters that best reproduce the non-parametric reconstruction (which is assumed to be a faithful representation of the underlying data). However, if the target model --- in this case, the Cauchy distribution --- is a poor fit to the data, then the non-parametric samples $q$ are highly unlikely to reproduce its functional form.
Each sample from the non-parametric reconstruction is mapped to a set of Cauchy parameters that minimise the metric defined in Eq.~\eqref{eq:metric}, but these samples are weighted by how well the original non-parametric fit matched the data, not by how well the distance-minimizing Cauchy model fitted the data. The 
two methods are therefore not expected to yield consistent results. The flexible binning procedure performs somewhat better, as it includes weighting by the quality of conversion factor. It is important to emphasise, however, that a large mismatch occurs only when the target model is a poor description of the data, in which case it should probably not be used and the posterior on its parameters carries little meaning. Moreover, this method has an in-built diagnostic to flag when this situation arises, in the form of the regularised concentration parameter, $\beta$. A low value of $\beta$ serves as an alarm bell to flag the considered model as potentially inadequate. 

\begin{figure}
    \centering
    \includegraphics[width=0.98\columnwidth]{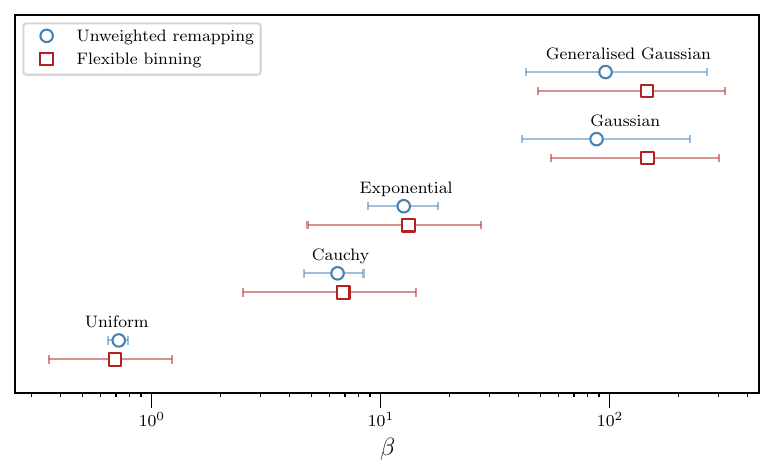}
    \caption{Summary of the $\beta$ values obtained for different target models when remapping from a flexible fit to data generated from a Gaussian distribution. Midpoints and bar ends show the median values and $90\%$ symmetric credible intervals of the $\beta$ posteriors for each model.}
    \label{fig:comp_beta}
\end{figure}

\subsection{Power-law+Peak}\label{sec:plp}
In order to illustrate the applicability of our approach to more complex 
populations, that are of astrophysical interest, we now present an example built on the \textsc{Power-law+Peak} model\footnote{Details of the \textsc{Power-law+Peak} model are reviewed in appendix~\ref{app:pop_model}.} used in \cite{KAGRA:2021duu}. 
Since the focus of these simulations is to demonstrate the remapping procedure and measurement errors and selection effects enter only in the initial non-parametric reconstruction, we have opted for simplicity to not include them in this example. Therefore, we assume a perfect measurement of all the events generated from the underlying distribution.

\subsubsection{GWTC-3 best fit parameters}\label{sec:gwtc3_data}
In this first example, we assume a \textsc{Power-law+Peak} model corresponding to the maximum-likelihood parameters\footnote{Values are given in appendix~\ref{app:pop_model}.} in the data release of GWTC-3 \cite{data_gwtc3_pop}, considering three different values for the number of observations, $N_o = 100$, $3,000$, and $10,000$. The prior on the hyperparameters is given in appendix~\ref{app:pop_model}.
We highlight that the observations considered here should not be directly compared to those reported in the GWTC-3 catalogue, as we have not included selection effects. 
More massive BBHs, such as the ones drawn from the Gaussian component of the \textsc{Power-law+Peak} model, have a higher detection probability and therefore, among the 69 observed events, a fraction larger than $1\%$\footnote{The fraction $\lambda$ of the Gaussian component for the maximum-likelihood parameters is $\lambda\sim0.01$.} is expected to come from the $35\ \mathrm{M}_\odot$ peak, allowing the feature to be easily resolved: in contrast, since we do not account for selection effects in our simulations, with 100 total events we expect only $\sim 1$ to be drawn from the Gaussian component.

\begin{figure}
    \includegraphics[width=0.98\columnwidth]{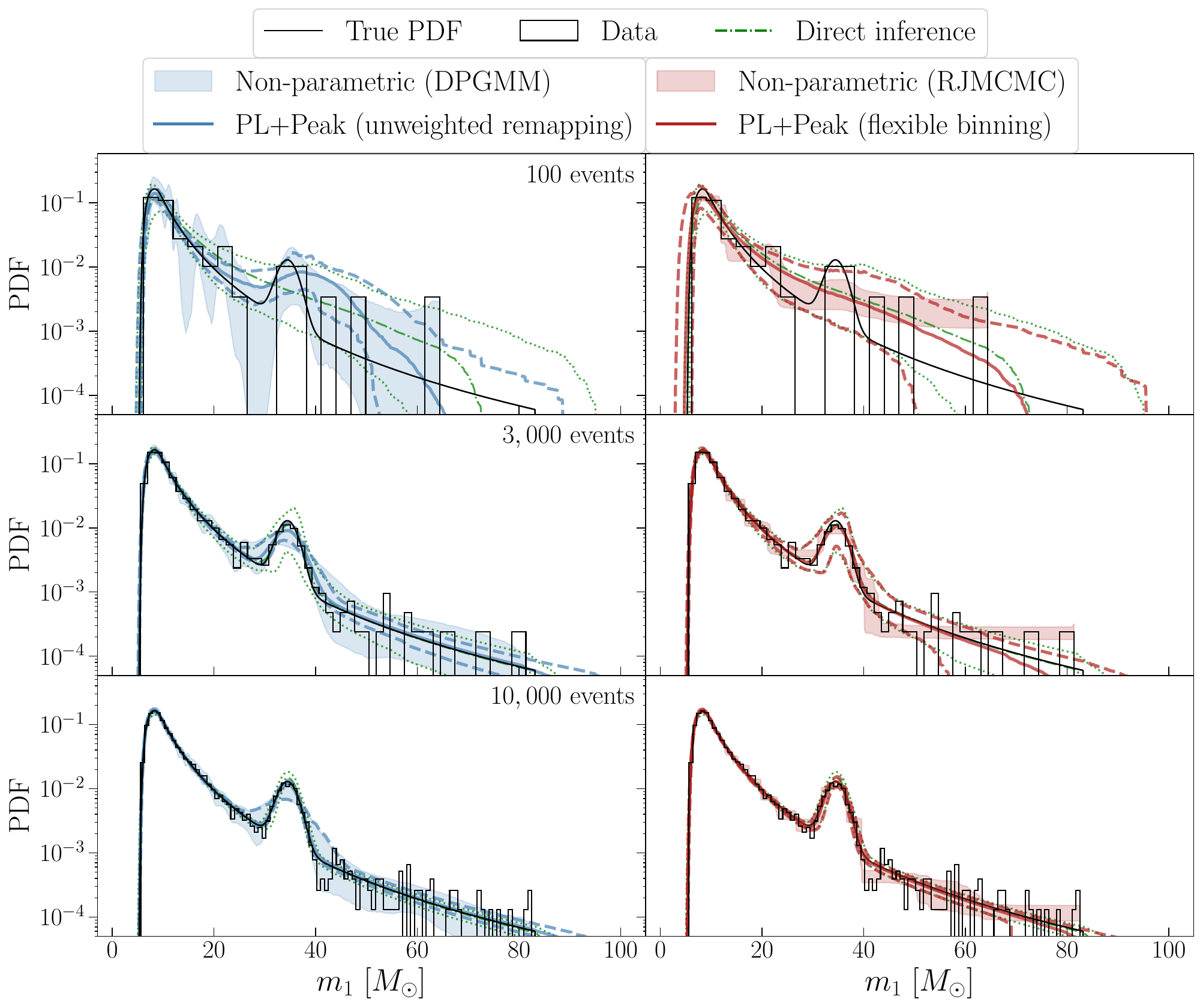}
    \caption{Comparison of the non-parametric reconstructions (shaded areas, 90\% credible region), the remapped \textsc{Power-law+Peak} distributions (dashed lines, median and 90\% credible region) and the true probability density function (solid black line), alongside with the histogram of the simulated data and the result obtained by direct inference (dot-dashed green lines, median and 90\% credible region). The left column refer to the unweighted remapping approach, the right one to the flexible binning method.}\label{fig:comp_plp}
\end{figure}

In figure~\ref{fig:comp_plp} we show an example of our remapping procedure applied to three datasets that differ in the number of observed events. For $N_o = 100$, the flexible binning approach does not unambiguously identify a peak and therefore the remapped \textsc{Power-law+Peak} posterior does not recover it beyond doubt, similarly to the direct inference case. The DPGMM reconstruction used in the unweighted remapping method is more sensitive to fluctuations in the data and hints at the presence of additional substructures, which is also reflected in the corresponding remapped \textsc{Power-law+Peak} posterior. As the number of events increases, the uncertainties from the non-parametric reconstructions decrease, and the remapped posteriors converge toward the distribution used to generate the data. The $90\%$ credible region obtained with the unweighted remapping approach remains broader, as its underlying non-parametric reconstruction yields larger uncertainties.
\review{In this case, the remapping procedure took about 5 minutes for the unweighted remapping and $o(10)$ minutes for the flexible binning approach regardless of the number of events. Once again these timings do not include the time to build the non-parametric reconstruction in the first instance.}

\begin{figure}
    \includegraphics[width=0.98\columnwidth]{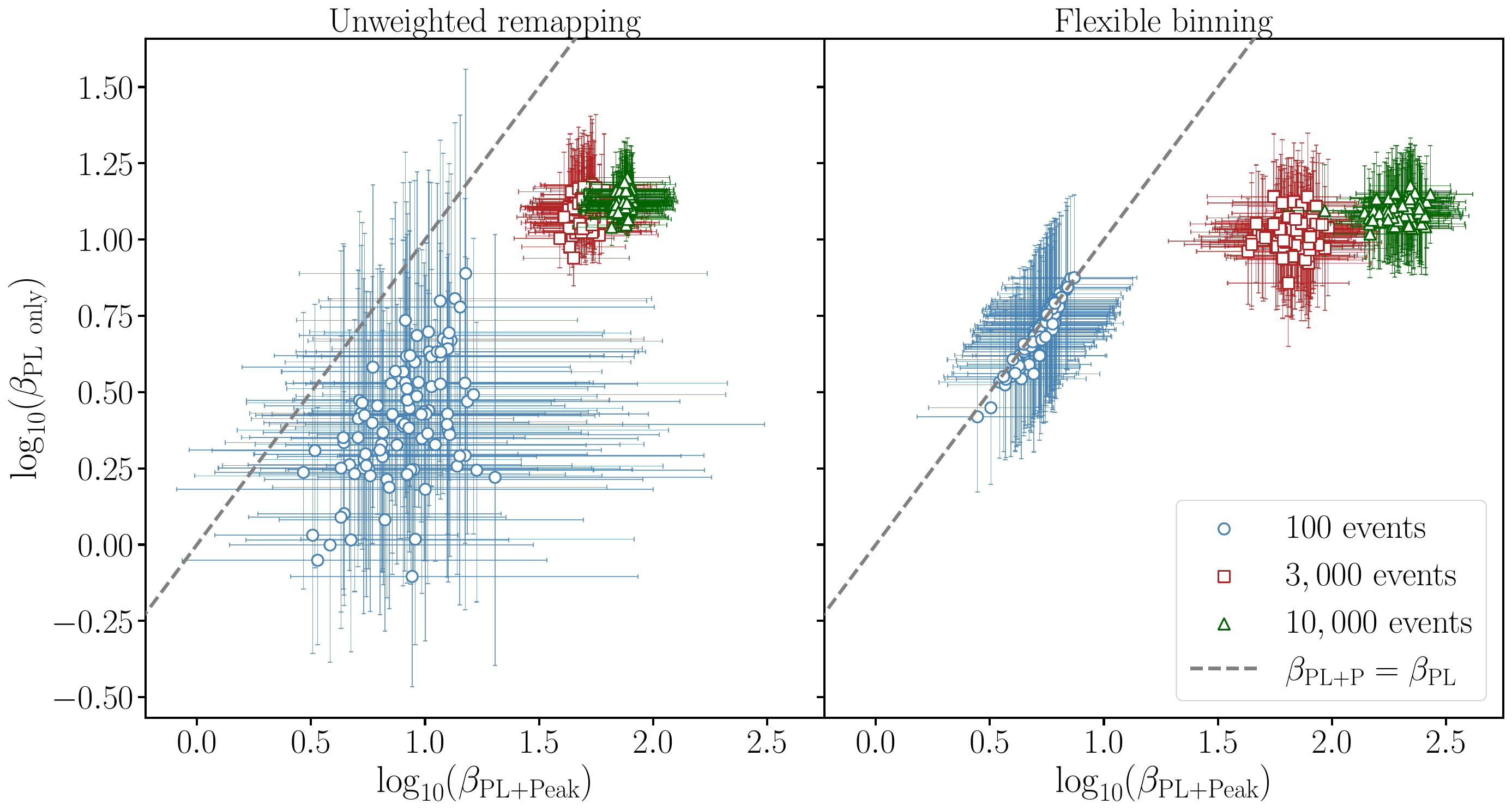}
    \caption{Comparison of the regularised concentration parameter when remapping to \textsc{Power-law+Peak} and to \textsc{Power-law} for different total numbers of observations. The error bars mark the $90\%$ credible interval.
    }\label{fig:comp_beta_plp}
\end{figure}

In order to illustrate the ability of the regularised concentration parameter $\beta$ to discriminate between models, we generate 100 datasets for each $N_o$, for a total of 300 datasets.
After having obtained a non-parametric reconstruction for each of these datasets, we apply the two implementations of our formalism using both the \textsc{Power-law+Peak} model (PL+Peak) and a \textsc{Power-law} model (PL only) in which the weight of the Gaussian component is set to zero.

Figure~\ref{fig:comp_beta_plp} compares the values of the regularised concentration parameter between the two models for each of the 300 available datasets. We observe a clear trend, with $\beta$ favouring the \textsc{Power-law+Peak} model more strongly as the number of events increases: for $N_o = 100$ there are barely any events in the Gaussian component and therefore we are not able to unambiguously assess the presence of the Gaussian feature. The degeneracy between the model is resolved as soon as more events are added to the dataset, with $\beta$ correctly pointing out the presence of the Gaussian feature in the simulated data.
The difference in the performance of the two approaches when applied to the most limited dataset arises from the effective prior over the space of probability density functions induced by the respective non-parametric models --- i.e., the data are not informative enough to properly constrain the underlying distribution and thus the specific details of the non-parametric reconstruction used play a more prominent role. This is related to the point highlighted at the beginning of this section while commenting on figure~\ref{fig:comp_plp}: the non-parametric method used in the flexible binning reconstruction does not unambiguously identify the presence of a peak at $\sim 35\ \mathrm{M}_{\odot}$, and therefore the remapping procedure gives similar $\beta$ values for the \textsc{Power-law} and \textsc{Power-law+Peak} models. 

This example illustrates that both approaches yield self-consistent results, while at the same time highlighting that the interpretation of the regularised concentration parameter is tied to the specific implementation used. For a chosen implementation, comparing $\beta$ values provides a valid measure of relative model performance: however, $\beta$ values obtained from different methods should only be directly compared as a broad ballpark estimate, since the numerical value depends on the specifics of the method --- namely, the non-parametric reconstruction used and the remapping technique.

\subsubsection{PP-plots}\label{sec:plp_prior}
To assess the statistical robustness of our approach, we now generate 100 realisations for each value of $N_o$, each time drawing the hyperparameters of the \textsc{Power-law+Peak} model from the prior distribution specified in appendix~\ref{app:pop_model}. For each of these realisations, we apply both implementations of our remapping procedure, using the same prior on the \textsc{Power-law+Peak} hyperparameters during inference.

\begin{figure}
    \includegraphics[width=0.98\columnwidth]{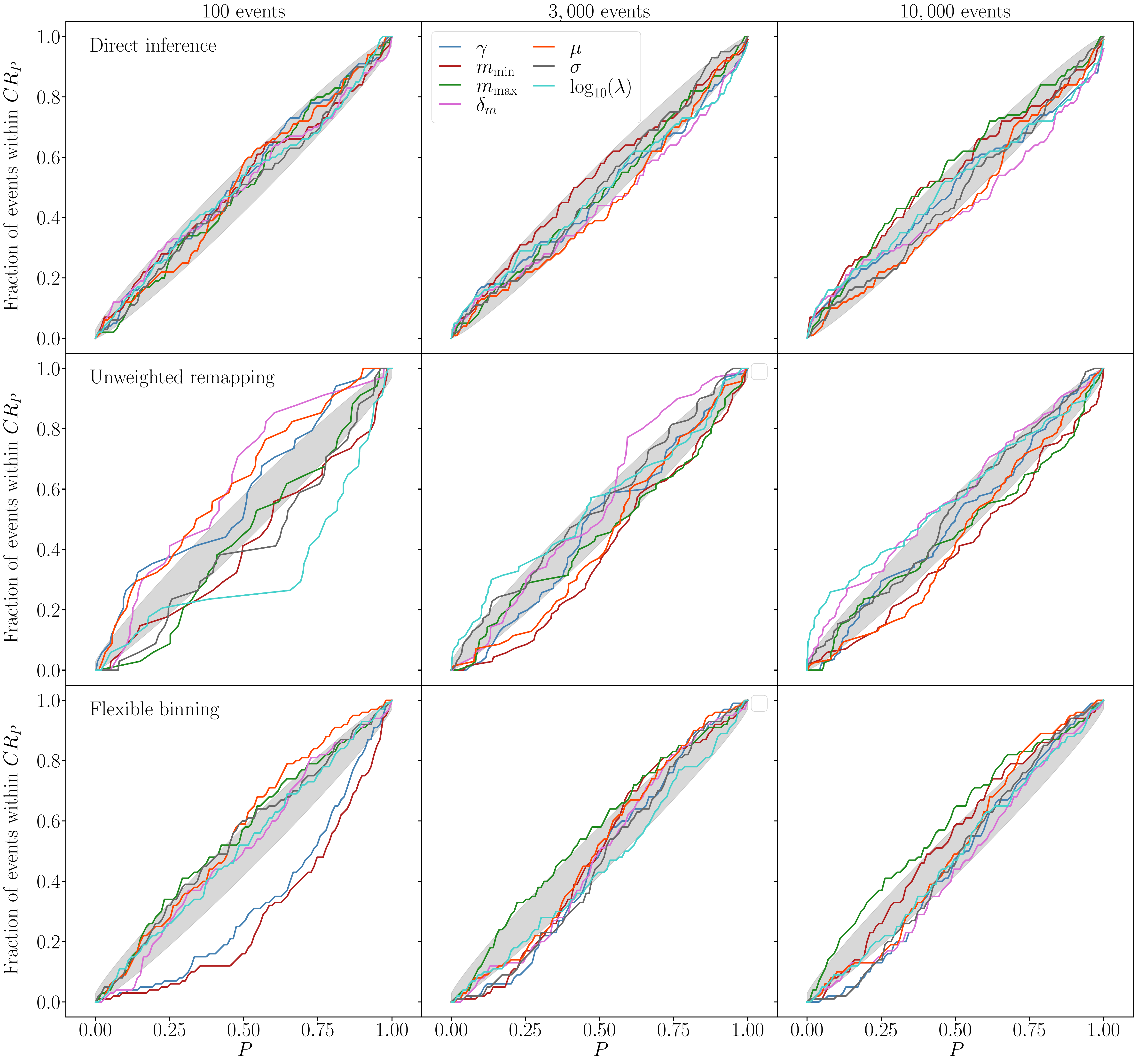}
    \caption{PP-plots: fraction of realisations for which the true parameter values are found within the $P\%$ confidence region ($CR_P$) as a function of the confidence level ($P$) for the different parameters of the \textsc{Power-law+Peak} model. The shaded gray area marks the $90\%$ credible interval estimated according to \cite{Cameron:2010bh}.
    Each column corresponds to a different number of events (left to right: $100$, $3000$ and $10000$ events) and each row to a different method (top to bottom: direct inference, unweighted remapping and flexible binning).
    }\label{fig:pp_plot}
\end{figure}

\begin{figure}
    \includegraphics[width=0.98\columnwidth]{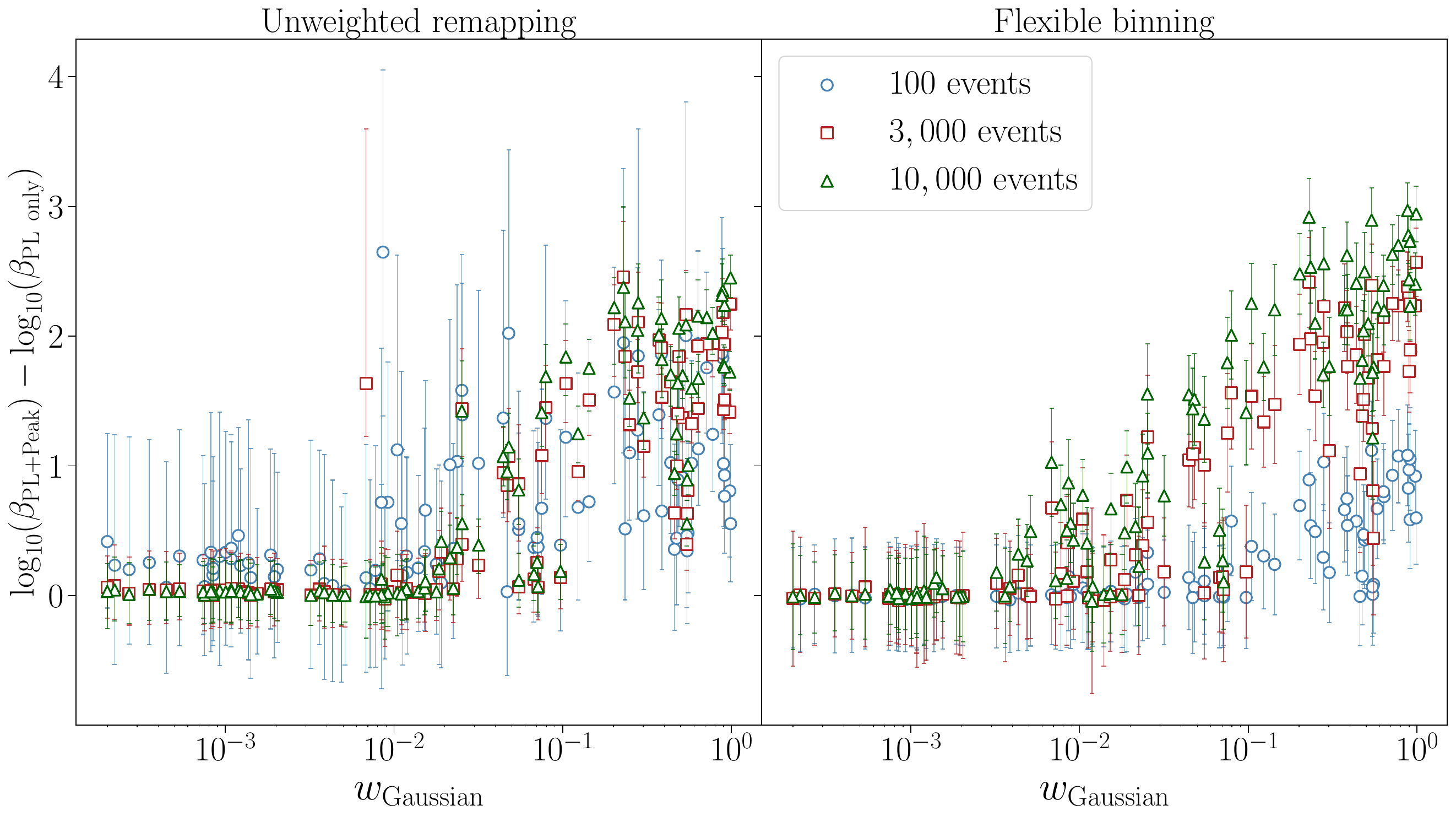}
    \caption{Comparison of the regularised concentration parameter when remapping to \textsc{Power-law+Peak} and to \textsc{Power-law} for different total number of observations. The error bars mark the $90\%$ credible interval.
    }\label{fig:comp_beta_plp_prior}
\end{figure}

The resulting PP-plots for both methods are shown in figure~\ref{fig:pp_plot}, together with the ones obtained by performing direct inference on the same data. The credible regions are estimated following \cite{Cameron:2010bh}. Since the model used to generate the data differs from the one employed in the analysis, perfect diagonality in the PP-plots is not necessarily expected: nonetheless, we observe that as the number of events increases, the PP-plots become increasingly consistent with the uniform percentile distribution within the statistical uncertainties for most parameters, eventually resembling the ones obtained via direct inference. \review{This behaviour is due to the ability of the non-parametric method to approximate the underlying distribution with different numbers of available observations.}

\review{For a very small number of observations $N_o$, the reconstruction is not expected to closely resemble the underlying distribution due to the insufficient amount of information carried by the limited sample pool.}
\review{More specifically, the uncertainty associated with the reconstruction would be such that, when marginalising over the posterior on the hyperparameters of the non-parametric model, the confidence region will indeed encompass the true distribution in the region where data are available, but due to the finite number of samples there is no reason to expect that any of the individual non-parametric curves resembles the underlying distribution (e.g., in this case a \textsc{Power-law+Peak}). Therefore, even if the target distribution belongs to the same family as the true one, the remapping procedure could still yield biased results.}
\review{As $N_o$ increases, the reconstruction better captures the main features of the true distribution, albeit still with substantial uncertainties. In this regime, the remapping procedure produces results consistent with the true values but with broader uncertainties than those from direct inference (see, e.g., figure~\ref{fig:comp_post}). While each individual inference may still be consistent with the true values of the hyperparameters, the presence of broad uncertainties can result in non-diagonal PP-plots.}
\review{For sufficiently large $N_o$, the non-parametric reconstruction accurately represents the true distribution and thus the remapping procedure yields results that are, in principle, identical to those from direct inference (see section~\ref{sec:dirichlet_process}) and, in practice, nearly indistinguishable -- therefore leading to diagonal PP-plots, within the statistical uncertainty.}

As discussed in section~\ref{sec:dirichlet_process}, if the underlying distribution is in the target model family (as is the case here), our remapping procedure will converge to the correct hyperparameters in the limit of a large number of events, provided that the family of models used for the initial inference on the data is sufficiently flexible that it can embed the true underlying distribution. This condition is expected to hold approximately true under the assumption that the non-parametric reconstruction is sufficiently flexible. 

Finally, we illustrate the sensitivity of our method to distinguishing the \textsc{Power-law+Peak} model from the \textsc{Power-law} model as a function of the prominence of the Gaussian component.
We define $w_{\rm Gaussian}$ as the total weight of the Gaussian component, i.e., the integral of the Gaussian distribution (accounting for the low-mass smoothing function) in the total probability density function:
\begin{equation}
   w_{\rm Gaussian}=\frac{\int \lambda S(m_1) {\rm G}(m_1) {\rm d}m_1}{\int S(m_1) \Big [ \lambda {\rm G}(m_1) +(1-\lambda) {\rm PL}(m_1)  \Big ] {\rm d}m_1}\,.
\end{equation}
The functions are defined in appendix~\ref{app:pop_model}.
Figure~\ref{fig:comp_beta_plp_prior} reports the difference in $\log_{10}(\beta)$ between the two models as a function of $w_{\rm Gaussian}$ for the three values of $N_o$ considered in this section. As the contribution of the Gaussian component increases, the \textsc{Power-law+Peak} model is increasingly favoured over the \textsc{Power-law} model. This preference becomes stronger with a larger number of events, consistent with our previous findings.

\section{Conclusions}\label{sec:conclusions}
As the computational cost of hierarchical Bayesian analyses increases more than linearly with the number of observations, there is a growing need for efficient methods to compare different models to the available data: in this work, we presented a new formalism to address this challenge. The method proposed in this paper is based on a two-step approach and involves performing one single initial non-parametric reconstruction incorporating all computationally intensive aspects, such as selection effects and measurement errors. This reconstruction, used as a form of data compression, is then remapped during the second step onto the model of interest --- ultimately producing a posterior distribution on the parameters of that model --- through an approach with reduced implementation complexity and computational costs \review{that are already absorbed in the first, reusable step}. We demonstrated that this procedure yields unbiased results and illustrated its application in the reconstruction of a population of astrophysical interest inspired by the ones currently employed by the LVK collaboration in the analysis of BBH population. 

Crucially, our model depends on the non-parametric reconstruction accurately representing the data. For models that exhibit sufficient flexibility such as the ones used in this work, this condition is more robustly met as the number of observations increases. Notably, this is the very regime where computational costs escalate, making the remapping approach all the more timely. Additionally, this method provides a self-consistent absolute measure of the goodness of fit for the model onto which the data is being remapped, offering a straightforward criterion for comparing different models. This goodness-of-fit measure, unlike the Bayes factor, can be evaluated even for models without free hyperparameters such as the populations predicted by astrophysical simulations, enabling a quantitative basis for comparison between such models.

In this work, we have focused on remapping between normalised distributions using the formalism of Dirichlet processes. However, the method is highly general and can also be applied to unnormalised functions --- such as the differential rate --- assuming any stochastic process that generates functions defined in the relevant space (e.g., strictly positive functions). Such extensions, as well as applying the remapping approach presented here to other contexts in which agnostic population studies are relevant (e.g., tests of General Relativity), will be the subject of future works.

\acknowledgments
The authors are grateful to Gregorio~Carullo, Walter~Del~Pozzo, Cecilia~Maria~Fabbri, Davide~Gerosa and to the organisers and participants of the IFPU Focus Week ``Emerging methods in GW population inference'' for discussions and comments.

This work was funded by the Deut\-sche For\-schungs\-ge\-mein\-schaft (DFG, German Research Foundation) – project number 546677095. S.R. acknowledges financial support from the European Research Council for the ERC Consolidator grant DEMOBLACK, under contract no. 770017, and from the German Excellence Strategy via the Heidelberg Cluster of Excellence (EXC 2181 - 390900948) STRUCTURES. The authors acknowledge support by the state of Baden-W\"urttemberg through bwHPC and the German Research Foundation (DFG) through grants INST 35/1597-1 FUGG and INST 35/1503-1 FUGG. A.T. is supported by MUR Young Researchers Grant No. SOE2024-0000125, ERC Starting Grant No.~945155--GWmining, Cariplo Foundation Grant No.~2021-0555, MUR PRIN Grant No.~2022-Z9X4XS, MUR Grant ``Progetto Dipartimenti di Eccellenza 2023-2027'' (BiCoQ), and the ICSC National Research Centre funded by NextGenerationEU.

\clearpage
{\centering\textbf{\huge Appendices}
\vspace{0.1in}}
\appendix

\appendix
\section{Poisson process}\label{app:poisson}
In section~\ref{sec:dirichlet_process}, we derived the map between $q$ and $p$ using a Dirichlet process. In this appendix, we present an analogous derivation in the case in which the functions we are dealing with are differential rates $q_R(\theta)$ and $p_R(\theta|\Lambda)$: in this case, the stochastic process will be a Poisson process. Here we will assume a binning scheme $\bar{B}$. Given a binning scheme, the differential rate $q_R(\theta)$ induces a differential rate in bin $i$ of
\begin{equation}
    \bar{Q}_{R,i} = N_t\bar{Q}_i \qq{with} \sum_{i=1}^{N_b} \bar{Q}_i = 1\, \qq{and} N_t = \int q_R(\theta) \, {\rm d}\theta,
\end{equation}
and equivalently for $\bar{P}_{R,i}$. 
We additionally assume that the total rate, $N_t$, is equal for the two models (the extension to unequal rates is straightforward). Finally, we interpret $Q_{R,i}$ as the number of events observed in bin $i$ and assume that the number of counts in each bin is independent and distributed following a Poisson distribution with rate $\bar{P}_{R,i}$:
\begin{equation}
    p(\bar{Q}_R|\bar{P}_R) = \prod_{i=1}^{N_b} \frac{(N_t\bar{P}_i)^{N_t\bar{Q}_i} e^{-N_t \bar{P}_i}}{\Gamma(N_t\bar{Q}_i)}\,.
\end{equation}
Making use of the Stirling approximation for the Gamma function, we can rewrite this expression as
\begin{equation}
    p(\bar{Q}_R|\bar{P}_R)\simeq \left(\prod_{i=1}^{N_b}\sqrt{\frac{N_t\bar{Q}_i}{2\pi}}\right)\prod_{i=1}^{N_b}\qty(\frac{N_t^{\bar{Q}_i}\bar{P}_i^{\bar{Q}_i}e^{-\bar{P}_i}}{N_t^{\bar{Q}_i}\bar{Q}_i^{\bar{Q}_i}e^{-\bar{Q}_i}})^{N_t} \,,
\end{equation}
or equivalently
\begin{equation}
    p(\bar{Q}_R|\bar{P}_R)\simeq \qty(\frac{N_t}{2\pi})^{\frac{N_b}{2}}\exp\qty[\qty(\frac{1}{2}\sum_{i=1}^{N_b} \ln(\bar{Q}_i) - N_t\sum_{i=1}^{N_b} \bar{Q}_i\ln\qty(\frac{\bar{Q}_i}{\bar{P}_i}) + N_t\sum_{i=1}^{N_b} \bar{Q}_i - N_t\sum_{i=1}^{N_b} \bar{P}_i)]\,.
\end{equation}
Making use of the normalisation condition on $\bar{P}$ and $\bar{Q}$, this becomes
\begin{multline}
    p(\bar{Q}_R|\bar{P}_R)\simeq \qty(\frac{N_t}{2\pi})^{\frac{N_b}{2}}\exp\qty[\qty(\frac{N_b}{2}\int \ln(q(\theta))\dd\theta -\frac{N_b}{2}\ln(N_t N_b) - \int q(\theta)\ln\qty(\frac{q(\theta)}{p(\theta|\Lambda)})\dd \theta)]\\ \hspace{-2.5cm}=\qty(\frac{N_t}{2\pi})^{\frac{N_b}{2}}\exp\qty[\qty(\frac{N_b}{2}\int \ln(q(\theta))\dd\theta - \frac{N_b}{2}\ln(N_t N_b) - D_{KL}(p||q))] \,,
\end{multline}
where the first integral term is finite for sufficiently regular functions and the second integral term is the KL divergence. In this case, we see that the map induced by the Poisson process has a simple form and corresponds to associating to each $q$ the value of $\Lambda$ that minimises the KL divergence between $q(\theta)$ and $p(\theta|\Lambda)$. Allowing for differences in the total rate between the target and initial model leads to a similar expression, with an extra tunable parameter representing the ratio of the rates.

\section{Population models}\label{app:pop_model}
In this appendix, we specify the functional forms for the families used in section~\ref{sec:demonstration}. Here $\mathcal{U}[a,b]$ indicates a uniform distribution in the range $a \leq u \leq b$.
\subsection{Section~\ref{sec:toy_model}: 
Gaussian distribution}

\underline{Gaussian distribution}:
\begin{equation}
    G(x, \mu, \sigma) = \frac{\exp(-\frac{(x-\mu)^2}{2\sigma^2})}{\sqrt{2\pi}\sigma}\,,
\end{equation}
Priors:
\begin{itemize}
 \item $\mu: \mathcal{U}[-1,2]$
 \item $\sigma: \mathcal{U}[0.1, 1.5]$
\end{itemize}

\ 

\noindent\underline{Generalised Gaussian distribution}:
\begin{equation}
    GenG(x, \mu, a, b) = \frac{b\,\exp[-\qty(\frac{|x-\mu|}{a})^\beta]}{2a\,\exp\qty(\Gamma(1/b))}\,,
\end{equation}
Priors:
\begin{itemize}
 \item $\mu: \mathcal{U}[-2,3]$
 \item $a: \mathcal{U}[0.1, 5]$
 \item $b: \mathcal{U}[1,4]$
\end{itemize}

\ 

\noindent\underline{Exponential distribution}:
\begin{equation}
    E(x, x_0,\lambda) = \frac{\exp(\frac{-|x-x_0|}{\lambda})}{2\lambda}\,,
\end{equation}
Priors:
\begin{itemize}
 \item $x_0: \mathcal{U}[-3,2]$
 \item $\lambda: \mathcal{U}[0.1, 5]$
\end{itemize}

\ 

\noindent\underline{Cauchy distribution}:
\begin{equation}
    C(x, x_0,\gamma) = \frac{1}{\pi\gamma}\frac{1}{1+\qty(\frac{x-x_0}{\gamma})^2}\,,
\end{equation}
Priors:
\begin{itemize}
 \item $x_0: \mathcal{U}[-3,2]$
 \item $\gamma: \mathcal{U}[0.1, 10]$
\end{itemize}

\ 

\noindent\underline{Uniform distribution}:
\begin{equation}
    \mathcal{U}(x): \frac{1}{x_{max}-x_{min}}
\end{equation}
\subsection{Section~\ref{sec:plp}: Power-law+Peak}
The fiducial model for the primary mass ($m_1$) distribution in the population analysis of GWTC-3~\cite{KAGRA:2021duu} is the \textsc{Power-law+Peak} model, defined by
\begin{equation}
    PLPeak(m_1,\Lambda)=S(m_1,m_{min},\delta_m) \Big [ \lambda {\rm G}(m_1,\mu,\sigma) +(1-\lambda) {\rm PL}(m_1,m_{min},m_{max},\gamma)  \Big ]\,, \label{eq:ext_pl_gauss} 
\end{equation}
 where ${\rm PL}$ is a power-law between $m_{\rm min}$ and $m_{\rm max}$ with slope $-\gamma$,
\begin{equation}
{\rm PL}(m_1,m_{min},m_{max},\gamma) =
	\begin{cases}
		\mathcal{N} m_1^{-\gamma} \qif m_{min} \leq m_1 \leq m_{max} \\ 
		\qq{0} \;\,\qotherwise
	\end{cases}\,, \label{eq:pl}
\end{equation}
with $\mathcal{N}$ being the appropriate normalisation factor, and $S(m_1,m_{\min},\delta_m)$ is the smoothing function introduced in \cite{Talbot:2018cva}: 
\begin{equation}
S(m_1,m_{min},\delta_m) =
	\begin{cases}
		0 \qif m_1 < m_{min}  \\
  [f(m_1-m_{min},\delta_m)+1]^{-1} \qif m_{min} \leq m_1 \leq m_{min}+\delta_m \\ 
		1 \qif m_1 > m_{min}+\delta_m \label{eq:smooth}
	\end{cases},
\end{equation} 
with $\delta_m$ defining the scale over which the $m_1$ probability density function goes smoothly to zero and
\begin{equation}
    f(m', \delta_m) = \exp\qty(\frac{\delta_m}{m'} + \frac{\delta_m}{m'-\delta_m})\,.
\end{equation}
Based on the GWTC-3 data release \cite{data_gwtc3_pop}, the maximum-likelihood parameters, used to generate the data in section~\ref{sec:gwtc3_data}, are:
\begin{itemize}
    \item $\lambda=0.019$,
    \item $\mu=34.5 M_{\odot}$,
    \item $\sigma=1.9 M_{\odot}$,
    \item $\gamma=3.5$,
    \item $m_{\rm min}=4.8 M_{\odot}$,
    \item $m_{\rm max}=83.1 M_{\odot}$,
    \item $\delta_m=5.5 M_{\odot}$.
\end{itemize}
We use the following prior on the parameters of the \textsc{Power-law+Peak} model: 
\begin{itemize}
    \item $\log_{10}(\lambda): \mathcal{U}[-4,0]$,
    \item $\mu: \mathcal{U}[20M_{\odot},50M_{\odot}]$,
    \item $\sigma: \mathcal{U}[1M_{\odot},10M_{\odot}]$,
    \item $\gamma: \mathcal{U}[1.1,10]$,
    \item $m_{\rm min}=\mathcal{U}[2M_{\odot},10M_{\odot}]$,
    \item $m_{\rm max}: \mathcal{U}[30M_{\odot},100M_{\odot}]$,
    \item $\delta_m: \mathcal{U}[0.5M_{\odot},10M_{\odot}]$.
\end{itemize}
This is the same prior from which we draw the parameters of the model to simulate events in section~\ref{sec:plp_prior}. 
For the \textsc{Power-law} model, we use the same prior on $\gamma$, $m_{\rm min}$, $m_{\rm max}$ and $\delta_m$. 

\bibliographystyle{JHEP}
\bibliography{bibliography}

\end{document}